\documentstyle[10pt,ctagsplt,righttag,amstext,amsfonts,amstex]{amsart}

 \newtheorem{thm}{Theorem}

 \theoremstyle{definition}
 
 \theoremstyle{remark}
 \newtheorem{rem}[thm]{Remark}
 \numberwithin{equation}{section}

 \newcommand{ \Min}{\operatornamewithlimits{Min}}
 \newcommand{ \E}{\operatorname{E}}
 \newcommand{ \var}{\operatorname{Var}}
 \newcommand{ \cov}{\operatorname{Cov}}
 \newcommand{ \corr}{\operatorname{Corr}}
 
 \newcommand{\beq}{\begin{equation}}
 \newcommand{\eeq}{\end{equation}}
\begin{document}

\title[Control Variates for  exotic options ]
 {Some Control Variates for  exotic options }

\author{ JC  Ndogmo}

\address{
\begin{flushright}
\em Department of Mathematics\\
University of the Western Cape\\
P/B X17, Bellville 7535\\[2mm]
Email: {\tt \bf jndogmo@@uwc.ac.za}
\end{flushright}
}

\begin{abstract}
There are no known exact formulas for the valuation of a number of
exotic options, and this is particularly true for options under
discrete monitoring and for American style options.  Therefore, one
usually recourses to a Monte Carlo Simulation approach, amongst
other numerical methods, to estimate the value of these options. The
problem which then arises with this method is one of variance
reduction. Control variates are often used, and we present some
results concerning these control variables, for the valuation of
Asian and lookback options. An inequality on functions of
correlations useful for comparing estimators in variance reduction
procedures is also provided.

\end{abstract}

\maketitle

\section{The problem of exotic options valuations}
There are two main types of options. Standard European or American
put and call options are referred to as vanilla options. These are
exchange traded options and their payoffs depend on the price of the
underlying asset at exercise time only. Contrary to these, there are
derivatives with more complicated payoffs and which are referred to
as exotic options.  Most exotic options trade in over-the-counter
markets and have been designed to meet the particular needs of
customers (corporations).

One important class of exotic options consists of path-dependent (or
history-dependent) derivatives. They have a payoff that depends not
just on the final price of the underlying asset, but also on the
path followed by this price. Asian options and lookback options are
typical examples of path-dependent options. In some instances,
numerical methods are the only means available to value these exotic
options, and in addition to Monte Carlo methods, other commonly used
numerical methods include the finite difference and the finite
element methods, the trinomial tree method, and some times the
Markov Chain method (see \cite{{MC, MCh, tt, barraq, mayo}} and the
references therein). Until the last decade, Monte Carlo method has
been considered as costly and unreliable, but based on innovative
techniques which are now a topic of current research, they've
yielded more promising results. In this context, Barraquand and
Martineau \cite{barraq}, amongst others, obtained some interesting
results.\par

Consider an option on a given security having $n$ days to maturity.
Let $S_d(j)$ denote the closing price of the security at the end of
day $j\, (\text{ for }j=1,\dots, n ).$ Assume also that the risk
free interest rate is $r$ per year,  and denote by $N$ the number of
trading days in a year. With these notation, the payoff from an
asian option is either given by

\begin{equation} \label{eq:1} Y= e^{-rn/N}\left( S_d(n) - \sum_{j=1}^n
\frac{s_d(j)}{n} \right)^+,
\end{equation}
where the strike price $K$ is taken as the average of all prices up
to the expiry date, or

\begin{equation} \label{eq:2}
Y= e^{-rn/N} \left(
\sum_{j=1}^n \frac{S_d(j)}{n}-K \right)^+,
\end{equation}

when the underlying's terminal price is taken as the average of all
prices up to maturity. The payoff from a lookback option is given by

\begin{equation} \label{eq:3}
Y= e^{-rn/N} \left( S_d(n) - \Min_{j=1,\dots,n} S_d(j) \right)^+.
\end{equation}

Barrier options are another type of frequently traded path-dependent
options. These are options for which a barrier value $v,$ which may
be a one- or two-dimensional vector, is specified, and the option
becomes alive or cease to exist according to the path followed by
the price w.r.t. the barrier value. When a barrier option is alive
at maturity, its payoff is that of the equivalent vanilla option,
i.e. that with the same strike price $K$, and number of days $n$ to
maturity.

The main problem with using Monte Carlo Simulation to value
path-dependent derivatives is that the computation time necessary
to achieve a reasonable level of accuracy can be excessively high.
Control variate techniques are some of the tools available for
implementing the necessary variance reduction procedures, which
can lead to dramatic savings in computation time.

Our focus in this paper will be on the valuation of Asian and
lookback options. In section 2 we discuss Monte Carlo valuation
simulations for these options and explain how controlled variates
techniques can be gainfully employed to improve the accuracy of the
valuation. In section 3, we present some results concerning the
control variables for an improvement of the variance reduction
procedure, and an inequality  on functions of correlation
coefficients useful for comparing estimators in variance reduction
techniques is also provided.

\section{Monte carlo simulations and variance reduction procedures}

Suppose that the price $S(t)$ of a given security follows a risk
neutral geometric Brownian motion with constant parameters $\mu$ and
$\sigma^2.$ This means in particular that for all $y \geq 0,$ and $t
\geq 0, \; \ln(S(t+y)/S(y))$ has a normal distribution with mean
$\mu t$ and standard deviation $\sigma \sqrt{t},$ and that $\mu = r-
\sigma^2 / 2,$ where $r$ represent the risk-free interest rate.

We denote as usual by $S(0)$ the initial price of the security,
and by $S_d(j)$ the price of the security at the end of day $j\,
(\text{ for }j=1,\dots, n),$ where $n$ is the number of days to
maturity. Set

\begin{equation}\label{eq:4}
 X(i)= \ln \left( \frac{ S_d(i) }{S_d(i-1)} \right), \quad
(i=1,\dots, n).
\end{equation}

The random variables $X(1),\dots, X(n)$ are independent and
identically distributed with mean $\mu / N$ and variance $\sigma^2
/N.$ That is, $X(i) \sim \Phi (\frac{\mu}{N}, \frac{\sigma^2}{N}).$
A straightforward calculation using ~\eqref{eq:4} shows that

\begin{align}
S_d(i)& =  S_d(i-1) e^{X(i)} \label{eq:5}  \\
      &= S(0)e^{X(1)+ \dots + X(i)} \label{eq:6}
\end{align}

Values of $X(1), \dots,X(n)$ can be generated by a computer and
Eq.~\eqref{eq:5}) or ~\eqref{eq:6} can be used to sample a random
path for the price $S$ of the security.

The valuation simulation for an asian option whose payoff is
$$Y= e^{-rn/N}\left( \sum_{j=1}^n \frac{S_d(j) }{n} - K \right)^+$$
as in Eq. ~\eqref{eq:2} can be implemented as follows

\begin{enumerate}
\item Generate $n$ values for the normal random
variables $X(1),\dots, X(n)$.
\item Calculate the end-of-day prices $S_d(i)= S(0) e^{X(1)+ \dots + X(n)},$ for $i=1,\dots, n.$
\item Calculate the payoff of the derivative $Y_j = e^{-rn/N}
\left( \sum \frac{S_d(i)}{n}  - K\right)^+,$ for the $j$-th
simulation run.
\item Compute the average mean $\bar{Y} = \frac{1}{R}\sum_{1}^{R}
Y_j,$ where $R$ is the total number of simulation runs, i.e. the
number of times that steps (1) through (3) are performed.
\end{enumerate}

The value of $\bar{Y}$ represents an estimate of the exact cost
$Y$ of the derivative. We have

\begin{equation} \E[\bar{Y}] = \E[Y] \end{equation}

 and

\begin{equation} \label{eq:7} \var (\bar{Y}) = \var (Y) / R. \end{equation}

The last equality, Eq. ~\eqref{eq:7}, shows that the accuracy of $
\bar{Y}$ improves with the number of simulations.

The simulation procedure is similar for lookback and for most of
the path-dependent derivatives. Monte carlo simulation presents
some advantages. They can be used to value path-dependent
derivatives as well as those derivatives whose payoff depend only
on the final value $S$ of the security. They can also be used to
value barrier options type securities, whose payoffs may occur
several times before maturity. More discussions on the advantages
of Monte Carlo simulation techniques can be found in ~\cite[P.\,
408]{hull}.

The drawbacks with Monte Carlo simulation is that the amount of
time necessary to achieve a reasonable accuracy can be
unacceptably high. As remedial measures, a number of variance
reduction techniques are available, and they can lead to dramatic
savings in computation time. These include the antithetic variable
technique, the importance sampling, the stratified sampling, and
the control variate technique. All these procedures and many
others are described in \cite{hull, clew, boyle}.

We now move on to explain how control variate can be used to
improve the simulation valuations of derivatives. Suppose that in
a general setup we want to estimate

\begin{equation} \label{eq:7'} M= \E [Y] \end{equation}

by simulation. Suppose furthermore that in the process of
generating the random variables that determine $Y$ we also find
another random variable $V$ whose mean value is known to be $\mu_v
= \E [V].$ Then rather than using the value of $Y$ as an
estimator, we can replace it with another value

\begin{equation} \label{eq:8} W= Y + c (V- \mu_v), \end{equation}

where $c$ is a constant to be specified. The value of $W$ is an
estimate for $M$ since $ \E [W]= E[Y]= M.$ We have

\begin{equation} \label{eq:9}
 \var (W) = \var (Y) + c^2 \var(V) + 2 c \cov(Y,V)
 \end{equation}

The best estimator is obtained by choosing $c$ to be the value that
makes $\var (W)$ as small as possible. Elementary calculus shows
that the value of $c$ that minimizes $\var (W)$ is given by

\begin{equation} \label{eq:10}
 c^* = - \frac{\cov (Y,V)}{\var (V)} \end{equation}

 substituting this value back into Eq. ~\eqref{eq:9} yields
 $$ \var (W) = \var (Y) - \frac{ \corr^2(Y,V)}{\var(V)}.  $$
 That is,
 \begin{align} \label{eq:11}
 \frac{\var(W)}{\var (Y)} & = 1- \corr^2 (Y,V) \\
 & = 1-r^2,
 \end{align}

where the quantity $r = \corr (Y,V)$ is the correlation coefficient
between $Y$ and $V.$ The number $r^2$ is often called the
coefficient of determination between $Y$ and $V.$ Eq. ~\eqref{eq:11}
shows that the variance reduction obtained with the control variate
$V$ is completely determined by $\corr (Y,V).$ More precisely, the
reduction obtained is $100 \% \corr^2 (Y,V)$ and consequently, the
higher the correlation between $Y$ and $V,$ the greater the variance
reduction obtained will be.

\subsection{Control variates}

Suppose that the payoff $Y$ from an option on a given security is of
any of the forms given by the Eqs. ~\eqref{eq:1}-\eqref{eq:3}. With
the notation of the previous section, set

\begin{equation} \label{eq:12} V= \sum_{i=1}^n \alpha_i X(i) \end{equation}

where $\alpha_i$  are some weights to be determined, and take

\begin{equation} \label{eq:13} W =Y+ c(V - \E(V)) \end{equation}

as the new estimator for $Y$.

We have

\begin{align*}
\E [V] & = \sum_{i=1}^n \alpha_i \E [X(i)] = \sum_{i=1}^n
          \frac{\alpha_i}{N}\left(  r - \frac{\sigma^2}{2}\right), \\
\intertext{so that} W & = Y + c\left[\sum_{i=1}^n \alpha_i X(i) -
\frac{\alpha_i}{N}(r - \frac{\sigma^2}{2}) \right] \\
& = Y + \sum_{i=1}^n \beta_i \left[X(i) - (r - \frac{\sigma^2)}{2}
/ N \right]
\end{align*}
where

\begin{equation} \label{eq:14} \beta_i = c\, \alpha_i. \end{equation}

On the other hand, we have

$$ \var (W) = \var (Y) + \sum_{i=1}^n \beta_i^2 \var (X(i)) + 2
\sum_{i=1}^n \beta_i \cov(Y, X(i)) $$

and it readily follows that the values of $\beta_i$ that minimize
$W$ are given by

\begin{equation} \label{eq:14p} \beta_i^* = - \frac{\cov (Y,X(i))}{\var
(X(i))} \end{equation}

In terms of these optimal values of the $\beta_i,$ the smallest
possible value for $\var (W)$ is given by

\begin{align}
\var (W) &= \var(Y) + \sum_i \frac{\cov^2 (Y, X(i))}{\var (X(i))}
+ 2\sum_i -\frac{\cov^2(Y, X(i))}{\var (X(i))}    \notag \\
\intertext{That is} \var (W)& = \var(Y) - \sum_i
\frac{\cov^2(Y,X(i) )}{\var (X(i))} \\
\intertext{Consequently,} \frac{\var (W)}{\var (Y)} &=
1-\sum_i\corr^2 (Y,X(i)) \label{eq:16}
\end{align}

We have thus proven the following result

\begin{thm} Let $V= \sum_i \alpha_i X(i),$ for some arbitrary
weights $\alpha_i$ and $for i=1,\dots, n.$ Let $W= Y+ c(V - \E[V])$
be an estimator of Y for some constant $c.$
\begin{description}
\item[(a)] The optimal variance reduction is achieved with this
control variable $V$ for the values of  $\alpha_i$ and $c$ such
that
$$ c\alpha_i =  - \frac{\cov (Y,X(i))}{\var (X(i))}, \qquad
(i=1,\dots,n)$$
\item[(b)] The optimal variance reduction thus obtained is given
by
$$ \frac{\var (W)}{\var (Y)} = 1-\sum \corr^2 (Y, X(i)) $$
\end{description}
\end{thm}

\begin{rem}
\begin{enumerate}$\mbox{}$
\item It follows from part (a) of the theorem that a control
variate of the form $V= \sum_i \alpha_i X(i)$ gives rise to an
optimal reduction if and only if $ \alpha_i = - \lambda
\frac{cov(Y, X(i))}{\var (X(i)}$ for all $i=1,\dots,n$ and for a
nonzero constant $\lambda.$
\item This theorem implies that for all $\alpha_1, \dots,\alpha_n$ and
for every\\
constant $c$
\begin{align*}
\frac{\var(W)}{\var(Y)} & \geq 1 - \sum \corr^2 (Y, X(i))\\
\intertext{where} W &= Y + c\left[ \sum_i \alpha_i X(i) - \alpha_i
(r - \sigma^2/2) \right]
\end{align*}
\end{enumerate}
\end{rem}

In ~\cite[P.\, 138]{ross}, the problem of finding the values of
$\alpha_i \,(\text{ for }i=1,\dots,n)$ for the best control
variate of the form $V= \sum_i \alpha_i X(i)$, and alternatively,
the problem of finding the values of $c_1,\dots, c_n$ for the best
estimator of $Y$ of the form
$$ Y + \sum_{i=1}^n c_i \left( X(i)-
\frac{r-\sigma^2/2}{N}\right)$$ are posed. Theorem 1 gives an
answer to this question, by determining the values of $\alpha_i$
and $c_i$ and by indicating precisely the optimal variance
reduction that can be achieved.

\begin{thm}
Let $Y$ be a random variable and let $X_1, \dots, X_n$ be n
independent random variables. Then for all numbers
$\alpha_1,\dots, \alpha_n$
$$ \corr^2 \left(  Y, \sum_i \alpha_i X_i\right) \leq \sum_i \corr^2
(Y,X_i). $$
\end{thm}

\begin{pf}
Let $V= \sum_i \alpha_i X_i$ and let $W= Y + c(V - \E [V])$ be an
estimator of $Y$ for some constant $c.$ For any fixed values of the
$\alpha_i$ the optimal variance reduction given by Eq.
~\eqref{eq:11} is
$$ \frac{\var (W)}{\var (Y)} = 1 - \corr^2(Y,V)$$
This is not smaller than the optimal variance reduction obtained for
all possible values of the $\alpha_i$ and $c$ given by Eq.
~\eqref{eq:16}. Consequently, $1-\sum_i \corr^2 (Y,X_i) \leq 1-
\corr^2 (Y, \sum_i \alpha_i X_i);$ that is, $\corr^2 (Y, \sum_i
\alpha_i X_i) \leq \sum_i \corr^2 (Y,X_i).$
\end{pf}

The result stipulated by this theorem is certainly very important,
but it is likely to be unknown. Indeed in statistics books that
present the most comprehensive material on the topic of
correlation coefficient between random variables, there is rarely
any discussion of a relationship between functions of correlations
(see \cite{ostle, rees}).

\subsection{Applications}
\begin{enumerate}

\item If $V = \sum_i \alpha_i X_i$ where the $\alpha_i$ are
constants
independent of $X_i,$ then no estimator of $Y$ of the form $ W =
Y+ \sum_i \alpha_i X_i - \E [V]$ using $V$ as control variable can
lead to an optimal variance reduction, since by Theorem 1, the
$\alpha_i$ must depend on $X_i.$

\item Let $V= \ln (s_d(n)/ S(0)).$ By the definition of the
$X(i)$, it readily follows that $V= \sum_i X(i)$. Thus by the
preceding remark, taking $V$ as control variable cannot give rise
to the best estimator.
\end{enumerate}

\section{Conclusion}

We have made use of the variance reduction formula ~\eqref{eq:11}
obtained for an estimator of the payoff $Y$, with a given control
variable $V$, to derive the form of the best possible control
variable of the form $V = \sum_i \alpha_i X (i)$, where the random
variables $X(i)$ are as in Eq. ~\eqref{eq:4}. We've also obtained an
upper bound for the correlation coefficient between an arbitrary
random variable $Y$ and an arbitrary linear combination of
independent random variables. This inequality can be useful in
comparing estimators in variance reduction procedures.

A generalization of this type of inequalities between functions of
correlation coefficients is desirable, to compare for instance an
estimator obtained  with a control variable of the form $V= \sum_i
\beta_i S_d(i)$, with that obtained using a control variable of the
form $\sum_i \alpha_i X(i)$. In this instance, one would need a
general relationship between functions of correlation coefficients
of the type $\corr^2 (Y, \sum_i \beta_i S_d(i))$ and $\corr^2 (Y,
\sum_i \beta_i  \ln ( S_d(i) / S_d(i-1) ) )$. As stated in this
paper, no result on such type of relationship seems to be available
in the literature. Finally, the results obtained in this paper
clearly apply to any random variable for which variance reduction is
required, and not only to the specific case of exotic options
considered in this paper.
%
%

\end{document}